\documentclass[letter,twocolumn]{jpsj2} %% for letters
\title{$d$-Wave Spin Density Wave phase in the Attractive Hubbard Model with Spin Polarization}

\author{Hiromasa \textsc{Tamaki}$^1$, Kazumasa \textsc{Miyake}$^{1}$, and Yoji \textsc{Ohashi}$^{2,3}$}

\inst{
$^1$Department of Materials Engineering Science, Graduate School of Engineering Science, Osaka University, Toyonaka, Osaka 560-8531, Japan\\
$^2$Faculty of Science and Technology, Keio University, Hiyoshi, Yokohama 223-8522, Japan\\
$^3$CREST(JST), 4-1-8 Honcho, Saitama 332-0012, Japan}

\abst{We investigate the possibility of unconventional spin density wave (SDW) in the attractive Hubbard model with finite spin polarization. We show that pairing and density fluctuations induce the transverse $d$-wave SDW near the half-filling. This novel SDW is related to the $d$-wave superfluidity induced by antiferromagnetic spin fluctuations, in the sense that they are connected with each other through Shiba's attraction-repulsion transformation. Our results predict the $d$-wave SDW in real systems, such as cold Fermi atom gases with population imbalance and compounds involving valence skipper elements.}

\kword{$d$-wave SDW, superfluidity, superconductivity, charge density wave, attraction-repulsion transformation, attractive Hubbard model, Pauli paramagnetic effect, FLEX approximation}

\begin{document}
\maketitle
%%%%%%%%%%%%%%%%%%%%%%%%%%%%%%%%%%%%%%%%%%%%%%%%%%%%%%%%%%%%%%%%%%%%%%%%%
%Introduction
%%%%%%%%%%%%%%%%%%%%%%%%%%%%%%%%%%%%%%%%%%%%%%%%%%%%%%%%%%%%%%%%%%%%%%%%%

It is well known that the Pauli's paramagnetic effect (PPE) strongly affects superconductivity. Under a strong magnetic field where the conventional BCS state no longer exists\cite{clogston62,maki64}, the Fulde-Ferrell-Larkin-Ovchinnikov (FFLO) state has been predicted to occur\cite{flude64,larkin65}, where the Cooper pairs have a finite center-of-mass momentum. The FFLO state has been recently reported in heavy fermion compound CeCoIn$_5$\cite{matsuda07,radovan03,bianchi03} and organic superconductors such as $\kappa$-(BEDT-TTF)$_2$Cu(NCS)$_2$\cite{singleton00} and $\lambda$-(BETS)$_2$FeCl$_4$\cite{uji06}.\par
Recently, the PPE has been also discussed in a superfluid $^6$Li Fermi gas, where spin polarization is realized by trapping atoms in two hyperfine states with different populations\cite{zwierlein06,partridge06}. In a cold Fermi gas, we can tune various physical parameters, such as pairing interaction and particle density. In addition, loading a Fermi gas on an optical lattice produced by the standing wave of laser light\cite{greiner02,chin06}, one can realize the attractive Hubbard model. Thus, cold Fermi gases are useful for the study of PPE in the wide parameter region of both continuum and lattice systems.
\par
The attractive Hubbard model (with a nearest-neighbor repulsion) is also applicable to compounds involving the valence skipper elements, such as Ba$_{1-x}$K$_x$BiO$_3$\cite{varma88,micnas90,taraph95}. Indeed, the phase diagram in the $x-T$ plane of Ba$_{1-x}$K$_x$BiO$_3$ has been nicely explained based on this model. Superconductivity in these materials appears by carrier doping into the charge density wave (CDW) state.\cite{pei90}
\par
In this paper, we investigate the PPE in the attractive Hubbard model. When the spin polarization is finite, we predict the $d$-wave spin density wave ($d$SDW) near the half-filling, where the SDW order parameter has a $d$-wave symmetry. To show this, we employ two approaches. The first approach uses Shiba's attraction-repulsion transformation\cite{shiba72}. In the second approach, we numerically determine the stable region of $d$SDW within the fluctuation exchange (FLEX) approximation\cite{bickers89}.

%%%%%%%%%%%%%%%%%%%%%%%%%%%%%%%%%%%%%%%%%%%%%%%%%%%%%%%%%%%%%%%%%%%%%%%%%
%repulsion-attraction transformation
%%%%%%%%%%%%%%%%%%%%%%%%%%%%%%%%%%%%%%%%%%%%%%%%%%%%%%%%%%%%%%%%%%%%%%%%%
We consider the attractive Hubbard model under an external magnetic field $h$ in the $z$-direction, given by
\begin{eqnarray}
H&=&-t\sum_{(i,j),\sigma}
\Bigl[c^{\dagger}_{i\sigma}c_{j,\sigma}+{\rm h.c.}
\Bigr]
-U\sum_{i}n_{i\uparrow}n_{i\downarrow}\nonumber\\
& &-\sum_{i,\sigma}
\Bigl[
\mu+h\sigma^z
\Bigr]
n_{i\sigma}.
\label{eq1}
\end{eqnarray}
Here, $c^\dagger_{i\sigma}$ is the creation operator of a fermion with spin-$\sigma$, $n_{i\sigma}\equiv c_{i\sigma}^\dagger c_{i\sigma}$, and $\sigma^z=\pm1$ as $\sigma=\uparrow,\downarrow$. $t$ describes the nearest-neighbor hopping, and the summation in the first term is taken over nearest-neighbor pairs. $U>0$ is the on-site pairing interaction, and $\mu$ is the chemical potential. In the case of two-component Fermi atom gas, $\mu=(\mu_\uparrow+\mu_\downarrow)/2$ and $h=(\mu_\uparrow-\mu_\downarrow)/2$, where $\mu_\sigma$ is the chemical potential of atoms with (pseudo)spin-$\sigma$.  
\par
To discuss the possibility of $d$SDW, we assume a bipartite lattice (where the lattice points can be divided into two sublattices). In this case, we can always transform (\ref{eq1}) into the repulsive Hubbard model by the particle-hole transformation\cite{shiba72},
\begin{eqnarray}
 c_{i\uparrow}^{\dagger}\rightarrow \tilde{c}^{\dagger}_{i\uparrow},~~~ 
 c_{i\downarrow}^{\dagger}\rightarrow \tilde{c}_{i\downarrow}e^{{\rm i}{\bf Q}\cdot{\bf R}_i}.
\label{eq2}
\end{eqnarray}
where ${\bf Q}$ is chosen so as to satisfy $\exp({\rm i}{\bf Q}\cdot{\bf R})=-1$ for any translation vector ${\bf R}$ between two sublattice sites. In particular, ${\bf Q}=(\pi/a,\pi/a)$ in the case of square lattice (where $a$ is the lattice constant). The resulting repulsive Hubbard model is given by
\begin{eqnarray}
\tilde{H}&=&
-t\sum_{(i,j),\sigma}
\Bigl[\tilde{c}^{\dagger}_{i\sigma}\tilde{c}_{j,\sigma}+{\rm h.c.}
\Bigr]
+U\sum_{i}\tilde{n}_{i\uparrow}\tilde{n}_{i\downarrow}\nonumber\\
& &-\sum_{i,\sigma}
\Big[h+\frac{U}{2}+\left(\mu+\frac{U}{2}\right)\sigma^z
\Bigr]\tilde{n}_{i\sigma}.
\label{eq3}
\end{eqnarray}
\par
%%%%%%%%%%%%%%%%%%%%%%%%%%%%%%%%%%%%%%%%%%%%%%%%%%%%%%%%%%%%%%%%%%%%%%%%%%%%%%
\begin{fulltable}[ht]
\caption{Correspondence of order parameters between attractive and repulsive Hubbard models. CDW: charge density wave. T-SDW: transverse spin density wave. L-SDW: longitudinal spin density wave. SF: superfluid phase. $\gamma_{\bf p}=\cos(p_xa)-\cos(p_ya)$ is the basis function of $d_{x^2-y^2}$-wave symmetry. $g_d$ is a coupling constant which induces $d$-wave SDW or $d$-wave SF. In this table, ${\bf q}$ is an arbitrary momentum. In the 2D square lattice, we take ${\bf Q}=(\pi/a,\pi/a)$.}
\label{t1}
\begin{tabular}{c|c}
\hline attractive Hubbard model & repulsive Hubbard model\\
\hline\hline 
 CDW,
$\phi^{\rm CDW}_{\bf q}=U\sum_{{\bf k},\sigma}\langle c^{\dagger}_{{\bf k+q}\sigma}c_{{\bf k}\sigma}\rangle$&
 L-SDW,
$\phi^{\rm LSDW}_{\bf q}=U\sum_{{\bf k},\sigma}\sigma^z\langle \tilde{c}^{\dagger}_{{\bf k+q}\sigma}\tilde{c}_{{\bf k}\sigma}\rangle $\\\hline
 $s$-wave SF,
$\phi^{s\rm{SF}}_{\bf q}=U\sum_{{\bf k},\sigma}\langle c^{\dagger}_{{\bf k}\uparrow}c^{\dagger}_{{\bf q-k}\downarrow}\rangle$&
 T-SDW,
$\phi^{\rm TSDW}_{\bf Q+q}=U\sum_{{\bf k},\sigma}\langle \tilde{c}^{\dagger}_{{\bf k+Q+q}\uparrow}\tilde{c}_{{\bf k}\downarrow}\rangle$\\\hline
 $d$-wave T-SDW,&$d$-wave SF\\
 $\phi^{d{\rm SDW}}_{{\bf p},{\bf Q+q}}=g_d\gamma_{\bf p}\sum_{{\bf k}}\langle c_{{\bf k}\uparrow}^{\dagger}
 c_{{\bf k-Q-q}\downarrow}\rangle\gamma_{\bf k}$&
 $\phi^{d\rm{SF}}_{{\bf p},{\bf q}}=g_d\gamma_{\bf p}\sum_{{\bf k}}\langle \tilde{c}_{{\bf k}\uparrow}^{\dagger}\tilde{c}_{{\bf q-k}\downarrow}^{\dagger}\rangle\gamma_{\bf k}$\\
\hline
\end{tabular}
\end{fulltable}
%%%%%%%%%%%%%%%%%%%%%%%%%%%%%%%%%%%%%%%%%%%%%%%%%%%%%%%%%%%%%%%%%%%%%%%%%%%%%%%
One finds from (\ref{eq1}) and (\ref{eq3}) that roles of $\mu$ and $h$ are exchanged between the two. This means that the magnetic (doping) phase diagram in the attractive Hubbard model is mapped onto the doping (magnetic) phase diagram in the repulsive Hubbard model. In this mapping, the order parameter in each phase is also transformed by the particle-hole transformation in (\ref{eq2}). We show some examples in Table \ref{t1}. 
\par

Since the discovery of high-$T_{\rm c}$ cuprates, it has been widely recognized that the ((quasi) two-dimensional) {\it repulsive} Hubbard model has the $d_{x^2-y^2}$-wave superconducting phase near the half-filling. This originates from a pairing interaction mediated by antiferromagnetic spin fluctuations\cite{bickers89,scalapino86,giamarchi91,jujo99,lichten00,yanase01}. Using this, we find from Table \ref{t1} that the transverse $d$SDW phase appears in the {\it attractive} Hubbard model when spin polarization is finite. The possibility of $d$SDW based on the attraction-repulsion transformation have been pointed out by Ho {\it et al.}\cite{ho09}
\par
The possibility of $d$SDW has been discussed in the context of hidden ordered state in ${\rm URu_2Si_2}$\cite{ikeda98}. In the mean field theory, the $d$SDW is known to be triggered by a nearest-neighbor repulsion. In the attractive Hubbard model, this interaction is considered to be induced by higher order processes mediated by pairing and density fluctuations.
\par
The $d$SDW state has various interesting properties. The (transverse) $d$SDW in the 2D square lattice is characterized by the complex order parameter $\phi^{d{\rm SDW}}_{{\bf p},{\bf Q+q}}=g_d\gamma_{\bf p}\sum_{{\bf k}}\langle c_{{\bf k}\uparrow}^{\dagger}c_{{\bf k-Q-q}\downarrow}\rangle\gamma_{\bf k}$, where $\gamma_{\bf p}=\cos(p_x a)-\cos(p_ya)$ and $g_d$ is the interaction for inducing $d$SDW. Since $\phi^{d{\rm SDW}}_{{\bf p},{\bf Q+q}}$ has nodes, the low-temperature specific heat behaves as $C\propto T^2$. Because of the momentum dependent order parameter, the $d$SDW state has no magnetic ordering. Instead, it is accompanied by alternating circular spin current\cite{ozaki92}, given by
\begin{equation}
J^l_{i,j}=\frac{{\rm i}t}{2}\sum_{\alpha\beta}\left(c^{\dagger}_{i\alpha}\sigma^l_{\alpha\beta}c_{j\beta}-c^{\dagger}_{j\alpha}\sigma^l_{\alpha\beta}c_{i\beta}\right),
\label{eq4}
\end{equation}
where $\sigma^l_{\alpha\beta}$ is the Pauli matrices, and $l=x,y,z$ describes the spin component flowing from the $j$-th site to $i$-th site. In the 2D square lattice, the $d$SDW state with ${\bf q}=0$ gives 
\begin{eqnarray}
\langle J^x_{i,j}+{\rm i}J^y_{i,j}\rangle&=&
{\rm i}t\phi_{d{\rm SDW}}
(\delta_{x_i,x_j\pm a}\delta_{y_i,y_j}-\delta_{x_i,x_j}\delta_{y_i,y_j\pm a})\nonumber\\
& &\times\exp[{\rm i}{\bf Q}\cdot{\bf R}_{i}].
\label{eq4b}
\end{eqnarray}
Here, ${\bf R}_{i}=(x_i,y_i)$, and we have simply written the order parameter as $\phi_{{\bf p},{\bf q}=0}^{d{\rm SDW}}=g_d\gamma_{\bf p}\phi_{d{\rm SDW}}$. We find from (\ref{eq4b}) that the argument of $\phi_{d{\rm SDW}}$ determines the spin component of the spin current.
\par

%%%%%%%%%%%%%%%%%%%%%%%%%%%%%%%%%%%%%%%%%%%%%%%%%%%%%%%%%%%%%%%%%%%%%%%%%%%%%%
%microscopic theory based on FLEX
%%%%%%%%%%%%%%%%%%%%%%%%%%%%%%%%%%%%%%%%%%%%%%%%%%%%%%%%%%%%%%%%%%%%%%%%%%%%%%

In the second approach, we determine the phase diagram of the attractive Hubbard model in the 2D square lattice. For the repulsive Hubbard model, the FLEX approximation\cite{bickers89} is known to reasonably describe the $d$-wave superconducting phase transition temperature $T_{\rm c}$. In this paper, we apply this approximation to the attractive case in (\ref{eq1}), to examine the stability of the $d$SDW state. The FLEX approximation has been recently used to determine $T_{\rm c}$ in the BCS-BEC crossover regime of the attractive Hubbard model without spin polarization\cite{tamaki08}.
\par
\begin{figure}[tb]
\begin{center}
\includegraphics[scale=0.45]{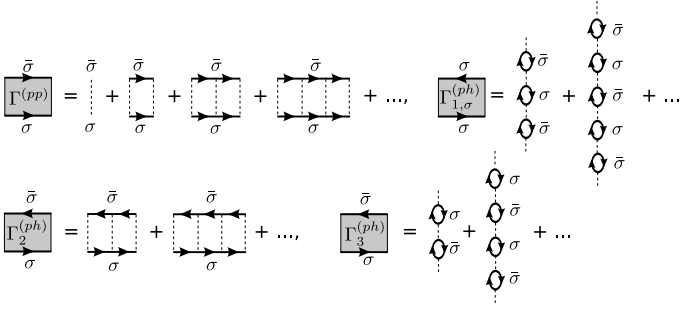}
\end{center}
\caption{Vertex functions of particle-particle channel $\Gamma^{\rm (pp)}$ and particle-hole channel  $\Gamma^{\rm (ph)}_{1,\sigma}$,  $\Gamma^{\rm (ph)}_2$, and  $\Gamma^{\rm (ph)}_3$, which are involved in the self-energy (\ref{eq:self}) and SDW Eliashberg equation (\ref{eq:eliash}). The solid and dashed lines describe $G_{\sigma}$ and U, respectively.}
\label{figd}
\end{figure}

In the FLEX approximation, the self-energy $\Sigma_\sigma({\bf k},{\rm i}\omega_n)$ in the single-particle Green's function $G^{-1}_{\sigma}({\bf k},{\rm i}\omega_m)={\rm i}\omega_m-\varepsilon_{\bf k}+\mu+h\sigma^z -\Sigma_{\sigma}({\bf k},{\rm i}\omega_m)$ (where $\varepsilon_{\bf k}=-2t[\cos(k_x a)+\cos(k_y a)]$) is given by
\begin{eqnarray}
\Sigma_{\sigma}({\bf k},{\rm i}\omega_m)
&=&T\sum_{{\bf q},n}\big[\Gamma^{\rm (pp)}({\bf q},{\rm i}\nu_n)G_{\bar\sigma}({\bf q-k},{\rm i}\nu_n-{\rm i}\omega_m)
\nonumber\\
& &+\Gamma^{\rm (ph)}_{1,\sigma}({\bf q},{\rm i}\nu_n)G_{\sigma}({\bf q+k},{\rm i}\nu_n+{\rm i}\omega_m)
\nonumber\\
& &+\Gamma^{\rm (ph)}_2({\bf q},{\rm i}\nu_n)G_{\bar\sigma}({\bf q+k},{\rm i}\nu_n+{\rm i}\omega_m)\big].
\label{eq:self}
\end{eqnarray}
Here, $\omega_m$ and $\nu_n$ are fermion and boson Matsubara frequencies, respectively, and $\{\bar{\uparrow},\bar{\downarrow}\}=\{\downarrow,\uparrow\}$. $\Gamma^{(pp)}$ and $\Gamma^{(ph)}_{j}$ $(j=1,2)$, respectively, describe the vertex functions in the particle-particle and particle-hole channels, having the forms
\begin{eqnarray}
 \Gamma^{\rm (pp)}({\bf q},{\rm i}\nu_n)
&=&\frac{-U}{1-U\Pi^{\rm (pp)}_{\uparrow\downarrow}({\rm q},{\rm i}\nu_n)},
\label{eq:g1}
\end{eqnarray}
\begin{eqnarray}
 \Gamma^{\rm (ph)}_{1,\sigma}({\bf q},{\rm i}\nu_n)
&=&\frac{U^4\Pi^{\rm (ph)}_{\sigma\sigma}({\bf q},{\rm i}\nu_n)\Pi^{{\rm (ph)}2}_{\bar\sigma\bar\sigma}({\bf q},{\rm i}\nu_n)}
{1-U^2\Pi^{\rm (ph)}_{\sigma\sigma}({\bf q},{\rm i}\nu_n)\Pi^{\rm (ph)}_{\bar\sigma\bar\sigma}({\bf q},{\rm i}\nu_n)},
\label{eq:g2}
\end{eqnarray}
\begin{eqnarray}
 \Gamma^{\rm (ph)}_2({\bf q},{\rm i}\nu_n)
&=&\frac{-U^3\Pi^{{\rm (ph)}2}_{\uparrow\downarrow}({\bf q},{\rm i}\nu_n)}
{1+U\Pi^{\rm (ph)}_{\uparrow\downarrow}({\bf q},{\rm i}\nu_n)}.
\label{eq:g3}
\end{eqnarray}
The diagrammatic structures of (\ref{eq:g1})-(\ref{eq:g3}) are shown in Fig.\ref{figd}
%%%%%%%%%%%%%%%%%%%%%%%%%%%%%%%%%%%%%%%%%%%%%
%%%%%%%%%%%%%%%%%%%%%%%%%%%%%%%%%%%%%%%%%%%%%%
The polarization functions in (\ref{eq:g1})-(\ref{eq:g3}) are given by
\begin{eqnarray}
\Pi^{\rm (pp)}_{\sigma\sigma'}({\bf q},{\rm i}\nu_n)=T\sum_{{\bf k},m}G_{\sigma}({\bf k},{\rm i}\omega_m)G_{\sigma'}({\bf q-k},{\rm i}\nu_n-i\omega_m),
\label{eq:p1}
\end{eqnarray}
\begin{eqnarray}
\Pi^{\rm (ph)}_{\sigma\sigma'}({\bf q},{\rm i}\nu_n)=-T\sum_{{\bf k},m}G_{\sigma}({\bf k},\omega_m)G_{\sigma'}({\bf q+k},{\rm i}\nu_n+i\omega_m).\label{eq:p2}
\end{eqnarray}
We briefly note that only $\Gamma^{\rm (pp)}$ involves the first and second order terms with respect to $U$ to avoid double counting. We calculate the particle density for each spin $n_{\sigma}$ from the condition $n_{\sigma}=T\sum_{{\bf k},m}G_{\sigma}({\bf k},{\rm i}\omega_m)e^{-{\rm i}\omega_m0^+}$.
\par
We determine the superfluid transition temperature $T_{\rm c}$ when the condition $\delta\equiv 1-U\Pi^{\rm (pp)}_{\uparrow\downarrow}({\bf q},0)=0.001$ is achieved. Although the criterion for $T_{\rm c}$ should be $\delta=0$, we take the finite value of $\delta$, because the convergence of (\ref{eq:self}) is not guaranteed when $\delta=0$ in the 2D system. We note that the superfluid (or FFLO) instability and CDW (or incommensurate CDW) instability occur simultaneously at the half-filling, because of the symmetry property of the model Hamiltonian in (\ref{eq1}). Since our formalism treats pairing and density fluctuations in a consistent manner, it satisfies this required condition. Indeed, one finds that $\Pi^{\rm (pp)}_{\uparrow\downarrow}({\bf q},0)=\Pi^{\rm (ph)}_{\uparrow\uparrow}({\bf Q+q},0)=\Pi^{\rm (ph)}_{\downarrow\downarrow}({\bf Q+q},0)$ at the half-filling.
\par
The SDW phase transition temperature is determined by solving the linearized Eliashberg equation in terms of the SDW order parameter. It is obtained from the anomalous self-energy in the Green's function $G_{\uparrow\downarrow}({\bf k},{\bf k+Q},\omega_m)\equiv -\int^{1/T}_{0}d\tau\langle T_\tau(c^{\dagger}_{{\bf k}_\uparrow}(\tau)c_{{\bf k+Q} \downarrow}(0)) \rangle e^{{\rm i}\omega_m\tau}$, as 
\begin{eqnarray}
\lambda\phi({\bf k},{\rm i}\omega_m)
&=&\sum_{{\bf k}',m'}V({\bf k},{\rm i}\omega_m;{\bf k}',{\rm i}\omega_{m'})
G_{\uparrow}({\bf k}',{\rm i}\omega_{m'})\nonumber\\
& &\times G_{\downarrow}({\bf k}'+{\bf Q},{\rm i}\omega_{m'})
\phi({\bf k},{\rm i}\omega_{m'}).
\label{eq:eliash}
\end{eqnarray}
Here, we have included both ${\bf k}$- and $\omega_m$-dependence of (transverse) SDW order parameter $\phi({\bf k},i\omega_m)$. The interaction $V({\bf k},i\epsilon_m;{\bf k}',i\epsilon_{m'})$ in the $d$SDW channel has the form
\begin{eqnarray}
V({\bf k},{\rm i}\omega_m;{\bf k}',{\rm i}\omega_{m'})&=&\Gamma^{\rm (pp)}({\bf k}'+{\bf k}+{\bf Q},{\rm i}\omega_{m'}+{\rm i}\omega_{m})\nonumber\\
& &+\Gamma^{\rm (ph)}_3({\bf k}'-{\bf k},{\rm i}\omega_{m'}-{\rm i}\omega_m),
\end{eqnarray}
\begin{eqnarray}
 \Gamma^{\rm (ph)}_3({\bf q},{\rm i}\nu_n)
&=&\frac{-U^3\Pi^{\rm (ph)}_{\uparrow\uparrow}({\bf q},{\rm i}\nu_n)\Pi^{\rm (ph)}_{\downarrow\downarrow}({\bf q},{\rm i}\nu_n)}
{1-U^2\Pi^{\rm (ph)}_{\uparrow\uparrow}({\bf q},{\rm i}\nu_n)\Pi^{\rm (ph)}_{\downarrow\downarrow}({\bf q},{\rm i}\nu_n)}.
\label{eq:eliash2}
\end{eqnarray}
The SDW phase transition is achieved when the maximum eigenvalue ($\equiv \lambda_{\rm max}$) in (\ref{eq:eliash}) reaches unity. To seek out $\lambda_{\rm max}$, we use a power method\cite{bickers89}. The resulting $\phi({\bf k},{\rm i}\omega_m)$ at the half-filling has the $d$-wave symmetry $\cos (k_{x}a)-\cos (k_ya)$, and is an even function with respect to $\omega_m$. In this regard, the odd-frequency SDW is not obtained unless the odd-frequency pairing is stabilized in the repulsive Hubbard model in (\ref{eq3}). Thus, when one considers a system obtained from the extended repulsive Hubbard model in which the odd-frequency pairing exists\cite{fuseya03,yada08} by the attraction-repulsion transformation, an odd-frequency SDW is expected to be realized.
\par
We briefly note that, since the spin rotational symmetry is broken when $h\ne 0$, the longitudinal $d$-wave spin susceptibility does not diverge, in contrast to the transverse one. As a result, the longitudinal $d$SDW is not obtained in the present model\cite{macridin04}.
\par
The $d$SDW vector ${\bf Q}$ deviates from ${\bf Q}=(\pi/a,\pi/a)$, when the carrier density is away from the half-filling. In addition, even-frequency SDW and odd-frequency SDW may mix with each other, because the non-diagnal matrix elements between even and odd frequency gap remain in the interaction $V({\bf k},{\rm i}\omega_m;{\bf k}',{\rm i}\omega_{m'})$ due to the particle-hole asymmetry of the density of states. However, in this paper, we only consider the $d$SDW order with ${\bf Q}=(\pi/a,\pi/a)$ and even frequency type for simplicity. In numerical calculations, we take $32\times 32$ lattice points and retain the fermion and boson Matsubara frequencies to $n_{\rm max}=m_{\rm max}=1024$.

%%%%%%%%%%%%%%%%%%%%%%%%%%%%%%%%%%%%%%%%%%%%%%%%%%%%%%%%%%%%%%%%%%%%%%%%%%%%%%%
%results
%%%%%%%%%%%%%%%%%%%%%%%%%%%%%%%%%%%%%%%%%%%%%%%%%%%%%%%%%%%%%%%%%%%%%%%%%%%%%%%
\begin{figure}[tb]
\begin{center}
\includegraphics[scale=0.45]{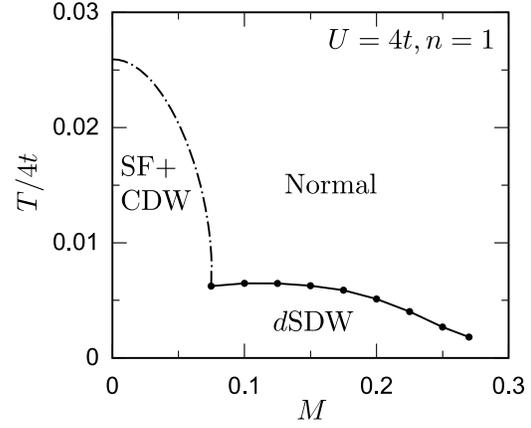}
\end{center}
\caption{Calculated phase diagram of the attractive Hubbard model in the 2D square lattice. The solid line with filled circle and dash-dotted line show $T_{\rm c}^{d{\rm SDW}}$ and $T_{\rm c}$, respectively. At the half-filling, $s$-wave superfluidity and CDW are degenerate below $T_{\rm c}$ even when $M=n_\uparrow-n_\downarrow$ is finite. Under the condition such as $U=4t$ and $n=1$, FFLO and Incommensurate CDW transition temperature is lower than $T_{\rm c}$ for ${\bf q}=0$ and $T_{\rm c}^{d{\rm SDW}}$.}
\label{fig1}
\end{figure}
\begin{figure}[tb]
\begin{center}
\includegraphics[scale=0.45]{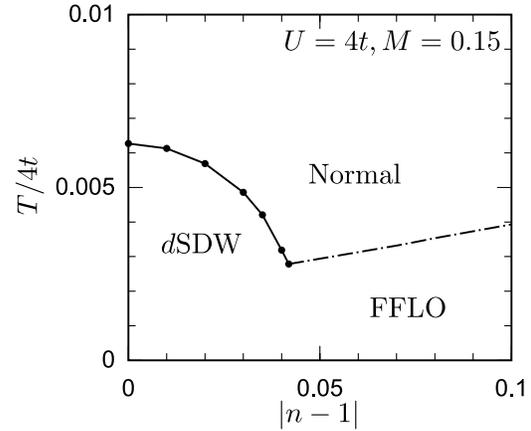}
\end{center}
\caption{Calculated $d$SDW phase transition temperature $T_{\rm c}^{d{\rm SDW}}$ and superfluid transition temperature $T_{\rm c}$ as a function of carrier doping. In this parameter region, FFLO phase appears with center-of-mass momentum ${\bf q}=(\pm3\pi/16a,0),(0,\pm3\pi/16a)$ at the $T_{\rm c}$.}
\label{fig2}
\end{figure}
\begin{figure}[tb]
\begin{center}
\includegraphics[scale=0.45]{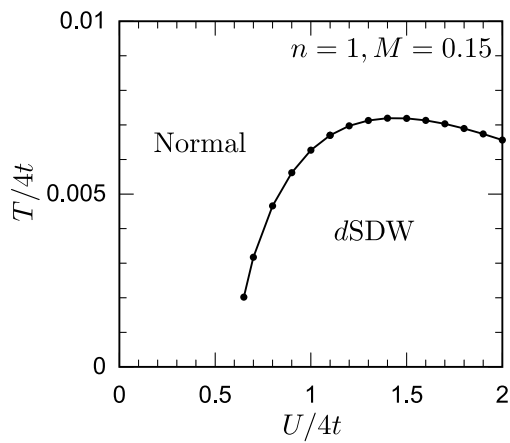}
\end{center}
\caption{Calculated $d$SDW phase transition temperature $T_{\rm c}^{d{\rm SDW}}$ as a function of pairing interaction $U$.}
\label{fig3}
\end{figure}
%%%%%%%%%%%%%%%%%%%%%%%%%%%%%%%%%%%%%%%%%%%%%%%%%%%%%%%%%%%%%%%%%%%%%%%%%%%%%%%

Figure \ref{fig1} shows the phase diagram of the attractive Hubbard model in the 2D square lattice. When the spin-polarization $M\equiv n_{\uparrow}-n_{\downarrow}$ is small, the $s$-wave superfluid phase appears. This phase corresponds to the antiferromagnetic phase in the repulsive Hubbard model. In the vicinity of this superfluid phase, we find the $d$SDW phase, corresponding to the $d$-wave superconducting phase in the repulsive case. The maximum $d$SDW phase transition temperature $T_{\rm c}^{d{\rm SDW}}$ equals $0.025t$.
\par
Fig. \ref{fig2} shows the doping dependence of $T_{\rm c}^{d{\rm SDW}}$ when $M=0.15$. The $d$SDW disappears at $|n-1|\sim 0.05$, but ${\bf q}=0$ superfluidity or FFLO superfluidity appears, depending on the $M$. This implies that the $d$SDW is not so robust against the carrier doping as the superfluid phase. Therefore, the symmetric density of states with respect to the Fermi level is favorable for the $d$SDW state. This $d$SDW-SF transition is discussed in the context of $d$SC-AF transition induced by the magnetic field in the repulsive Hubbard model\cite{yanase08}. We also note that the strong pairing interaction is crucial for the $d$SDW state, as shown in Fig. \ref{fig3} (although $T_{\rm c}^{d{\rm SDW}}$ again decreases when the interaction is very strong ($U\gg 4t$)).
\par
%%%%%%%%%%%%%%%%%%%%%%%%%%%%%%%%%%%%%%%%%%%%%%%%%%%%%%%%%%%%%%%%%%%%%%%%%%%%%%%
% summary and discussion
%%%%%%%%%%%%%%%%%%%%%%%%%%%%%%%%%%%%%%%%%%%%%%%%%%%%%%%%%%%%%%%%%%%%%%%%%%%%%%%

To conclude, we have discussed the possibility of $d$SDW state in the attractive Hubbard model with spin polarization. Then, we have predicted this novel SDW state based on the attraction-repulsion transformation, as well as numerical calculation within the FLEX approximation. 
\par
A cold Fermi gas loaded on an optical lattice is a strong candidate for the system to realize $d$SDW. In this paper, we considered a two-dimensional system in determining the phase diagram. From the previous work\cite{arita99,takimoto02} on the quasi-two-dimensional repulsive Hubbard model, one expects that the three-dimensionality suppresses the $d$SDW and hides it in the FFLO phase. Thus, the dimensionality of system should be close to two dimension for our purpose. In this regard, the optical lattice is very convenient, because the dimensionality can be tunable. For the observation of the $d$SDW state, the photoemission spectroscopy\cite{stewart08} would be useful, where the single-particle spectral weight is expected to show the $d_{x^2-y^2}$-wave gap structure. 
\par
The valence skipping compounds, ${\rm Ba_{1-x}K_xBiO_3}$\cite{pei90} and ${\rm BaPb_{1-x}Bi_xO_3}$, are also other candidates. Their phase diagrams are well described by the extended attractive Hubbard model involving the nearest neighbor repulsion. In the mean-field theory, this additional repulsion is known to be favorable for the stabilization of the $d$SDW state. Thus, it is a possible scenario that the $d$SDW appears in the vicinity of the CDW or superconducting phase under a magnetic field. 
\par
We note that the stability of $d$SDW against the orbital effect (which always exists in charged systems) is still unclear, although the $d$-wave density wave state is known to be insensitive to this\cite{nguyen02}. We also note that the $d$SDW may coexist with superfluid state or CDW near the phase boundary below $T_{\rm c}^{d{\rm SDW}}$. These are important future problems to understand the physics of unconventional spin density wave. 

%%%%%%%%%%%%%%%%%%%%%%%%%%%%%%%%%%%%%%%%%%%%%%%%%%%%%%%%%%%%%%%%%%%%%%%%%%%%%%%
%\section*{Acknowledgment}
%%%%%%%%%%%%%%%%%%%%%%%%%%%%%%%%%%%%%%%%%%%%%%%%%%%%%%%%%%%%%%%%%%%%%%%%%%%%%%%
One of us (K.M.) acknowledges Y. Fuseya for informing us that he independently reached a result on a possibility of $d$SDW in a quasi-one dimensional system. This work was supported by a Grant-in-Aid for Scientific Research (19340099, 19540420, and 20500044) from JSPS, and in part by a Grant-in-Aid for Specially Promoted Research (20001004) from MEXT of Japan. H.T was supported by Global-COE program (G10) from JSPS.

\end{document}